\documentclass[twocolumn,showpacs,preprintnumbers,amsmath,superscriptaddress,amssymb]{revtex4}

\usepackage{graphicx}
\usepackage{bm}

\begin{document}


\title{$\pi$-States in all-pnictide Josephson Junctions}

\author{C. Nappi}
\email{c.nappi@cib.na.cnr.it}
\affiliation{%
CNR Istituto di Cibernetica ''E. Caianiello'' Via Campi Flegrei 42
I-80078, Pozzuoli, Napoli, Italy}%

\author{S. De Nicola}
\affiliation{%
CNR Istituto Nazionale di Ottica I-80078, Via Campi Flegrei
42,Pozzuoli, Napoli, Italy}%
\affiliation{INFN Sezione di Napoli, Complesso Universitario di Monte S. Angelo, Via Cinthia, I-80126 Napoli, Italy}%

\author{M. Adamo}
\affiliation{%
CNR Istituto di Cibernetica ''E. Caianiello'' Via Campi Flegrei 42
I-80078, Pozzuoli, Napoli, Italy}%

\author{E. Sarnelli}
\email{e.sarnelli@cib.na.cnr.it}
\affiliation{%
CNR Istituto di Cibernetica ''E. Caianiello'' Via Campi Flegrei 42
I-80078, Pozzuoli, Napoli, Italy}%

\date{\today}

\begin{abstract}
We study the Josephson effect in $s_{\pm}/I/s_{\pm}$ junctions made by two bands reversed sign s-wave
($s_{\pm}$) superconductive materials. We derive an equation
providing the bound Andreev energy states  parameterized by
the band ratio $\alpha$, a parameter accounting for the weight of the second band with respect to the first
one at the interface. For selected values of the band ratio
and tunnel barrier amplitude, we
predict various features of the Josephson current, among which a possible high temperature $\pi$ state of
the junction (a doubly degenerate junction ground state) and a
$\pi \rightarrow 0$ crossover with decreasing temperature.
\end{abstract}

\maketitle

The importance of the Josephson effect as a tool to probe the
properties of new discovered superconductive materials can hardly be
underestimated. This has been true in the past in investigating the
d-wave cuprate materials and is nowadays the case for the multi-gap
iron based superconductors \cite{kamihara} whose complex behavior,
beyond the BCS theory, is presently a focus in condensed matter
theory \cite{mazin,chub} and material science \cite{blat}.

There is a rather solid indication now, supported by experiments,
that the pair potential symmetry in these compounds is s-wave with
sign-reversing order parameter ($s^\pm$). Several theoretical models
have already discussed the experimental consequences of an extended
s-wave ($s_{\pm}$- wave) order parameter symmetry on the Andreev
conductance of an $NS$ interface and on the Josephson effect in the
iron-based superconductor junctions \cite{parker,wu,ng,golubov,lind,
tsai, chen, ota, erin, berg, kosh,dag,lin,vak,apo}. In particular
the Josephson effect has been studied in hybrid devices i.e.
$sIs_{\pm}$ junctions, as summarized by Seidel in his review
\cite{seidel} and reference therein.

In this Letter we study an all-pnictide symmetric $s_{\pm}Is_{\pm}$ Josephson
junction and we discuss a number non-trivial physical
consequences on the Josephson
effect due due to the presence of a second conduction band.
We show that, depending on the band ratio parameter $\alpha$, which accounts for the
 weight the second band, a $\pi$-state \cite{bula}
can develop for a wide range of junction
transparencies and temperatures. This $\pi$-state can persist in the
full range of temperatures or can undergo a $\pi\rightarrow0$
(inverse $0\rightarrow\pi$) crossover as the temperature decreases,
depending on the band ratio parameter. This crossover is analogous
 to the $0\rightarrow\pi$ crossover \cite{ili,testa2},
found in mesoscopic d-wave superconductor Josephson  junction
\cite{tanaka}.

We adopt the simplest model of a Josephson junction that shows the
essential features of the Josephson effect in the presence of two
gaps, namely we consider a superconductor(S)-insulator(I)-superconductor(S) contact. The iron based junction
is modeled by considering a one-dimensional conductor, whose left
($x<0$) and right ($x>0$) halves are both two band metals (two
different states at the Fermi level, one with the wave vector $p$
and the other with $q$). We assume that the motion of quasiparticles
is described by the Bogoliubov de Gennes  (BdG) equation
\cite{degennes}
 and that the order parameter has already been obtained self-consistently from the gap equation.
 Therefore we choose a one dimensional model for the gap, such that the
left and right two band superconductors have pair potentials given
by
\begin{equation}
\label{eq:pair_potential}
 \Delta_j(x)=\Delta_je^{i \varphi_j}\theta(-x)+\Delta_je^{i
(\varphi_j+\varphi)}\theta(x), \quad j=1,2
\end{equation}
The normal region, where $\Delta_j=0$, has an infinitesimal width
and we also introduce a scattering potential $U(x)=U_0\delta(x)$.
The possibility of nodes in the gap function is not considered. In
the case of the two-gap model with unequal s-wave gaps, we write a
wave function of the same type introduced in the Blonder Tinkham
Klapwijk model \cite{BTK} as solution of the BdG equations and treat
the presence of the second band through the introduction of Bloch
wave functions \cite{golubov}. The bound state (B) eigenfunction
with energy $|E|<\Delta_1$ (we assume $\Delta_1<\Delta_2$) can be
written as
\begin{widetext}
\begin{eqnarray}
\label{eq:PSISd}
\Psi^{S_{L}}(x) =   \left(
\begin{array}{rl}
&u_B(x) \nonumber \\
 &v_B(x)
\end{array}
 \right)=a_B\left[\left(
\begin{array}{rl}
&u_1 \nonumber \\
 &v_1e^{-i \varphi_1}
\end{array}
 \right)\phi_{-p_e}(x)+\alpha_0\left(
\begin{array}{rl}
&u_2 \nonumber \\
 &v_2e^{-i \varphi_2}
\end{array}
 \right)\phi_{-q_e}(x)\right]+ \nonumber \\ b_B\left[\left(
\begin{array}{rl}
&v_1 \nonumber \\
 &u_1e^{-i \varphi_1}
\end{array}
 \right)\phi_{p_h}(x)+\alpha_0\left(
\begin{array}{rl}
&v_2 \nonumber \\
 &u_2e^{-i \varphi_2}
\end{array}
 \right)\phi_{q_h}(x)\right] \hspace{1cm}x<0
 \nonumber \\
 \nonumber \\
  \Psi^{S_{R}}(x) =  \left(
\begin{array}{rl}
&u_B(x) \nonumber \\
 &v_B(x)
\end{array}
 \right)= c_B\left[\left(
\begin{array}{rl}
&u_1 \nonumber \\
&v_1e^{-i( \varphi_1+\varphi)}
\end{array}
 \right)\phi_{p_e}(x)+\alpha_0\left(
\begin{array}{rl}
&u_2 \nonumber \\
&v_2e^{-i (\varphi_2+\varphi)}
\end{array}
 \right)\phi_{q_e}(x)\right]+ \nonumber \\ d_B\left[\left(
\begin{array}{rl}
&v_1\nonumber \\
 &u_1e^{-i (\varphi_1+\varphi)}
\end{array}
 \right)\phi_{-p_h}(x)+\alpha_0\left(
\begin{array}{rl}
&v_2 \nonumber \\
 &u_2e^{-i (\varphi_2+\varphi)}
\end{array}
 \right)\phi_{-q_h}(x)\right] \hspace{1cm}x>0
 \nonumber \\
\end{eqnarray}
\end{widetext}

where $\varphi$ is a global phase difference between the two superconductive regions and $\varphi_{1}$, $\varphi_{2}$ are
the phases of the gaps $\Delta_{1}$, $\Delta_{2}$ in both $p$ and $q$ bands respectively.
In the case of $s_{\pm}$ gap model, $\varphi_2-\varphi_1=\pi$.
In the considered energy range the wave function has to decay exponentially for $|x|\rightarrow \infty$. The coefficients $a_B,b_B,c_B,d_B$ are the probability amplitudes transmission with branch crossing or without branch crossing.
Essential above is the introduction of
$\alpha_0$,  a mixing coefficient defining the ratio of probability
amplitudes for a quasiparticle to be transmitted  to the first, ($p$),
or second, ($q$), band \cite{golubov}; the functions $\phi$'s are the Bloch waves in
the two-band superconductor; $p$ and $q$ are the Fermi vectors for
the two bands  corresponding to  the same energy $E$ \cite{golubov}:

\begin{eqnarray}
\phi_{\lambda}(x)=\sum_GC_{G,\lambda}\exp[i(\lambda+G)x]
\end{eqnarray}
where $\lambda=p_e,q_e,p_h,q_h$, and $G$ represent any reciprocal
lattice vector. $u_1$, $v_1$ and $u_2$, $v_2$ are the Bogoliubov
coefficients for the first and second  band, respectively
\begin{eqnarray}
&&u_1=\left[\frac{1}{2}\left(1+i\frac{\Omega_1}{E}\right)\right]^{1/2},
v_1=\left[\frac{1}{2}\left(1-i\frac{\Omega_1}{E}\right)\right]^{1/2}\nonumber \\
&&u_2=\left[\frac{1}{2}\left(1+i\frac{\Omega_2}{E}\right)\right]^{1/2},
v_2=\left[\frac{1}{2}\left(1-i\frac{\Omega_2}{E}\right)\right]^{1/2}\nonumber \\
&&\Omega_1=\sqrt{E^2-\Delta_1^2},\Omega_2=\sqrt{E^2-\Delta_2^2}
\end{eqnarray}

The global wave function $\Psi$ must satisfy the following boundary conditions at the interfaces $x=0$
 \begin{eqnarray}\label{eq:boundary}
&& \Psi^{S_L}(0^-)=\Psi^{S_R}(0^+) \nonumber \\
&&\Psi'^{S_R}(0^+)-\Psi'^{S_L}(0^-)= \frac{2mU_0}{\hslash ^2}\Psi^{S_R}(0^+)
 \end{eqnarray}
 where primes denote derivative with respect to $x$. \\
 Coupling between the
bands is implicit in the boundary condition  requirements. In fact, the assumed wave function $\Psi$ is of the form $\Psi=\Psi_1+\alpha_0\Psi_2$ where,
separately, $\Psi_1$ and $\Psi_2$ solve BdG equations for excitations of
energy $E$ for the gap $\Delta_1$ and $\Delta_2$, respectively,. The boundary conditions, Eqs. (\ref{eq:boundary}), are requested to be a constrain for the whole function $\Psi$.

The above matching procedure,  with the
  requirement of non-triviality of the solution for the coefficients $a_B,b_B,c_B,d_B$ provides for the $s_{\pm}/I/s_{\pm}$ junction, the spectral equation
\begin{eqnarray}
\label{eq:secular}
&&\left( -1+2\sqrt{1-E^2} \sqrt{r^2-E^2} \alpha^2-r^2 \alpha^4 \right)\left( 2Z^2+1\right)+ \nonumber \\
&&2E^2\left[Z^2(1+\alpha^4)+(1-\alpha^2+\alpha^4)\right]+ \nonumber \\
&&\left[-1+\alpha^2\left(\ 2E^2+2\sqrt{1-E^2}\sqrt{r^2-E^2}-r^2\alpha^2\right)\right] \times \nonumber \\ &&\cos(\varphi)=0
\end{eqnarray}
where we have introduced the gap ratio $r=\Delta_2/\Delta_1$, the band ratio parameter $\alpha=\alpha_0 \phi_q(0)/\phi_p(0)$ and the barrier strength $Z=U_0/\hbar v.$
The energy is given in units of $\Delta_1(T)$, and the
two gaps will be assumed to obey the same BCS-like temperature
law. Boundary conditions on the wave function derivatives are usually discussed in terms of
 Fermi velocities in the case of plane waves. For Bloch waves we have to introduce "interface velocities"
 \begin{eqnarray}
v_\lambda=-\frac{i \hbar}{m}\frac{\phi_\lambda'(0)}{\phi_\lambda(0)}
\end{eqnarray}
In deriving the spectral equation we have assumed, for the sake of simplicity, equal band interface velocities $v_{\lambda}=v$.

\begin{figure}[htbp]
\includegraphics[width=20pc]{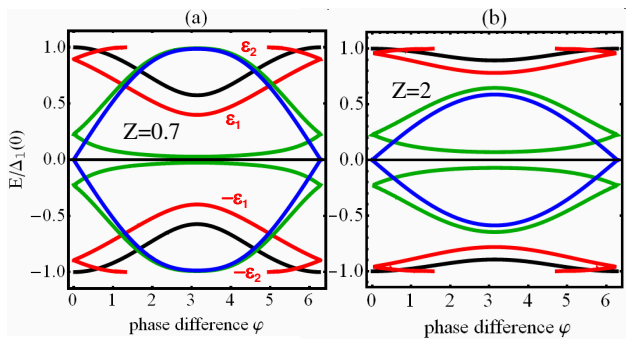}
 \caption{Andreev levels for a $s_{\pm}$
superconductor $S/I/S$ junction with $r=2$ and increasing values of
$\alpha$. (a): $Z=0.7$. Black line, $\alpha=0.01$, red line,
$\alpha=0.5$, green line, $\alpha=0.7$.  (b): $Z=2$. Black line,
$\alpha=0.01$, red line, $\alpha=0.4$, green line, $\alpha=0.7$. The
black line curves correspond to a nearly perfect ($\alpha=0$)
"conventional"  s-wave junction. The blue lines are the Andreev
level for $\alpha=\alpha_c=0.7071$. }\label{fig:iron1}
\end{figure}

An analysis of the spectral equation shows that for $\alpha\neq0$
there are, in general, four energy levels, $
E_1=\pm\epsilon_1(\varphi,\alpha)$ and $
E_2=\pm\epsilon_2(\varphi,\alpha)$ (see Fig. \ref{fig:iron1}). In a
single-band case ($\alpha=0$) Eq \ref{eq:secular} provides the well
known result for a conventional $SIS$ junction
  \cite{kulik,riedel}, namely
 $ E_1=\pm\epsilon_1(\varphi,0)=\pm \Delta_1(T) \left[{1-D \sin^2 (\varphi/2)}\right]^{1/2}$,
 where $D=1/(1+Z^2)$ is the transmission probability through the
$\delta$-function barrier, i.e. the junction transparency.

  As $\alpha$ increases, the energy levels $ E_2=\pm\epsilon_2(\varphi,\alpha)$
 start to
 branch off the levels $E_1=\pm\epsilon_1(\varphi,\alpha)$, while these latter gradually approach the zero energy state. As an example,
Fig. \ref{fig:iron1} (a) and (b) show the modifications of the
Andreev levels for increasing values of the band ratio, for two
values of the barrier parameter, $Z=0.7$ and $Z=2$ and for the band
gap ratio $r=2$, respectively. The condition for the existence of a
zero energy level can be easily obtained from the spectral equation.
It is given by
 $(1- r \alpha^2)^2(1 + 2 Z^2 + \cos(\varphi))=0$.
  Accordingly, a zero energy state $E_1=\pm \epsilon_1(\varphi,\alpha_c)= 0$,
  is obtained for the critical value of the band ratio
 $\alpha_c=1/\sqrt{r}$, for any $Z$ value. The  energy level $E_2=\pm \epsilon_2(\varphi,\alpha_c)$
corresponding to the critical value of the band ratio is given by \cite{note0}
\begin{eqnarray}\label{eq:limit}
&&E_2=\pm\epsilon_2(\varphi,\alpha_c)=\nonumber \\
&&\pm\sqrt{ \frac{
 2D r^2 (2 - D + D \cos \varphi) \sin^2(\varphi/2)}{1 + r (2 - 2 D + r) + 2 D r \cos \varphi}}
\end{eqnarray}

and it is shown as a blue line in Fig. \ref{fig:iron1} for the
indicated values of $D$ and $r$. For low transparencies ($D\ll1$),
Eq. \ref{eq:limit} reduces to $E_2=\pm
\epsilon_2(\varphi,\alpha_c)=\pm \Delta_1(T) 2r/(1+r)\sqrt{D} \sin
\left(\varphi/2\right)$, a result closely resembling the midgap
states of d-wave superconductor Josephson junction \cite{note}. For
$\alpha
> \alpha_c$, there are no surface bound states with real energy eigenvalues.

 In the nearly insulating limit ($Z\rightarrow \infty$), the system
decouples and we obtain information on the two separate electrodes. More precisely, in this
limit, the spectral equation (\ref{eq:secular}) describes surface bound states of energy
\begin{equation}
E_B=\pm\sqrt{\frac{1-r^2\alpha^4}{1-\alpha^4}}
\end{equation}
already discussed in ref. \cite{golubov} for a junction $N/I/s^\pm$.

For $0<\alpha <\alpha_c$, the two emerging levels $E_2=\pm
\epsilon_2 (\varphi,\alpha)$ and the levels $E_1=\pm \epsilon_1
(\varphi,\alpha)$ have, in general, opposite dispersions, i.e.
$dE_2/d\varphi>0$, $dE_1/d\varphi<0$ as can be seen in Fig.
\ref{fig:iron1}(a) and (b). The sign difference of the dispersions
is the key point in determining
  the Josephson current-phase relation, as it will be discussed below.

The Josephson current $I_d$ carried by the discrete Andreev levels
$E_k$ through the contact can be found from the free energy, according to
the following relation \cite{beenakker,riedel}
\begin{eqnarray}\label{eq:currlev}
I_d(\varphi,\alpha)=&&\frac{2e}{\hbar} \sum_{k=1,2}\frac{\partial
 E_k}{\partial \varphi}f(E_k)= \nonumber \\&& -\frac{2e}{\hbar}\sum_{k=1,2}\left(\frac{\partial
 \epsilon_k}{\partial \varphi}\tanh \frac{\epsilon_k}{2k_BT} \right)
\end{eqnarray}
where $f(E_k)$ is the Fermi distribution function and $k$ labels
the Andreev level with energy $E_k=\pm\epsilon_k(\varphi,\alpha)$.

\begin{figure}[htbp]
\includegraphics[width=20pc]{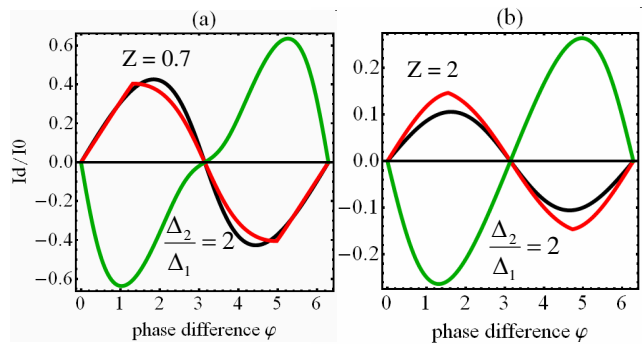}
 \caption{current-phase relation for a $s_{\pm}$
superconductor $S/I/S$ junction  with $r=2$, $T/T_c=0.01$ and
increasing values of $\alpha$.  (a)$Z=0.7$, (black
line,$\alpha=0.01$, red line, $\alpha=0.5$, green line,
$\alpha=0.7$). (b) $Z=2$. (black line, $\alpha=0.01$, red line,
$\alpha=0.4$, green line, $\alpha=0.7$)} \label{fig:iron2}
\end{figure}

\begin{figure}[htbp]
\includegraphics[width=22pc]{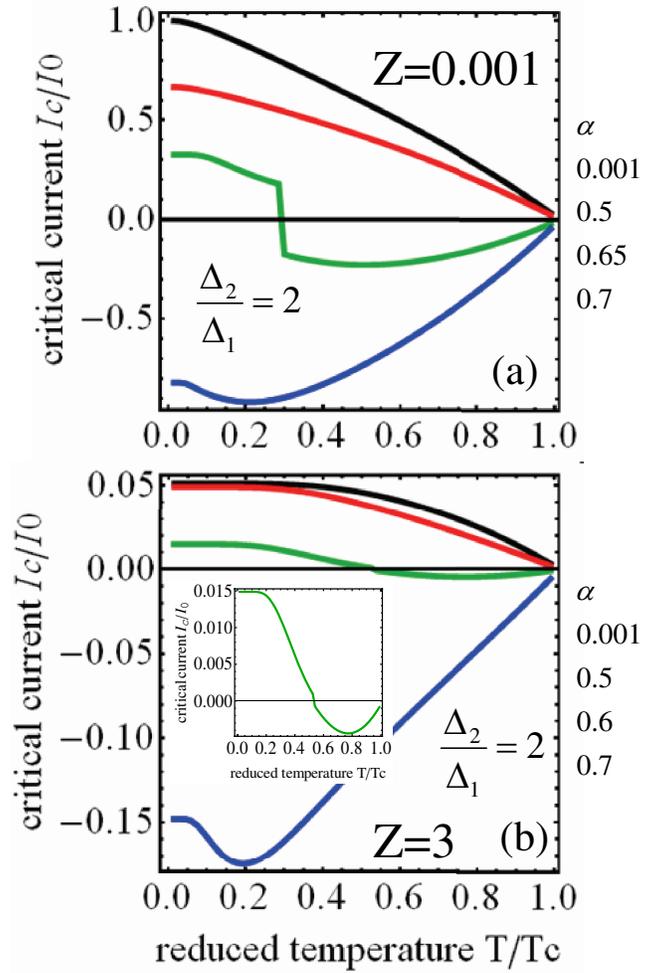}
 \caption{ Critical current as a function of
temperature for a $s_{\pm}$ superconductor $S/I/S$ junction  with
$r=2$, $T/T_c=0.01$ (a)$Z=0.001$; black line, $\alpha=0.001$, red line $\alpha=0.5$,
green line,$\alpha=0.65$, blue line, $\alpha=0.7$. (b) $Z=3$; black line, $\alpha=0.001$, red line, $\alpha=0.5$,
green line, $\alpha=0.6$, blue line, $\alpha=0.7$. The negative
sign of $I_c$ indicates that the junction minimum is at
$\varphi=\pi$ ($\pi$-junction). The inset in panel (b) shows, on a larger scale,
the transition $\pi\rightarrow0$ for $\alpha=0.6$}
\label{fig:iron3}
\end{figure}

Fig. \ref{fig:iron2}(a) and (b) show the current-phase relations
$I_d(\varphi,\alpha)$, at low temperature, corresponding to the
discrete spectrum of Andreev levels represented in Fig.
\ref{fig:iron1} (a) and (b) respectively and calculated through Eq.
(\ref{eq:currlev}). The current in this figure is normalized with
respect to $I_0=e \Delta_1(0)/\hbar$, which is the zero temperature
maximum Josephson current through a one dimensional s-wave junction
in the clean limit. The three curves correspond to values of the
band ratio approaching the critical value $\alpha_c=0.7071$. The
current-phase relation represented by the green lines ($\alpha=0.7$)
in both Figs \ref{fig:iron2} (a) and (b) shows a clear $\pi$
phase-shift (the maximum Josephson current for $0<\varphi<\pi$ has a
negative value). For the considered $\alpha$ value this $\pi$ state
persists in the whole temperature range (see Fig. \ref{fig:iron3}
(a) and (b)). The mechanism of formation of this state is the
following.  As the band ratio increases an upper
$+\epsilon_2(\varphi,\alpha)$ and a lower
$-\epsilon_2(\varphi,\alpha)$ extra Andreev bands gradually emerge.
These two bands have  different character compared to the low energy
bands $\pm\epsilon_1(\varphi,\alpha)$: they transport supercurrents
in the opposite directions. As the temperature decreases, only low
energy level are populated while those at higher energy are empty.
In the competition between these opposite carrying current energy
levels $\pm\epsilon_2(\varphi,\alpha)$ and
$\pm\epsilon_1(\varphi,\alpha)$, which coexist for any value of the
band ratio, it is the temperature that determines the direction of
the total current and the possible existence of a $\pi \rightarrow
0$ crossover in
 the considered two gap superconductor one-dimensional junction.

 The situation is similar, although not identical, to that
found in a two-dimensional d-wave $\pi$ junction. In this case
\cite{tanaka}, two kinds of bands, conventional and midgap,
alternate, without coexisting and compete each other in determining
the supercurrent, depending on the angle of incidence of the Andreev
quasiclassical trajectories with the interface. Therefore it is the
two-dimensionality here that plays the key role.

In Fig. (\ref{fig:iron3}), (a) and (b),  we report, for increasing
values of $\alpha$ , the dependencies of the normalized Josephson
critical current $I_c/I_0$ ($I_c =
\text{max}_{\varphi}I_d(\varphi)$) from the reduced temperature
$T/T_c$, derived by the discrete Andreev spectra for a nearly clean
junction ($Z=0.001$) and a low transparency junction ($Z=3$),
respectively.

\begin{figure}[htbp]
\includegraphics[width=20pc]{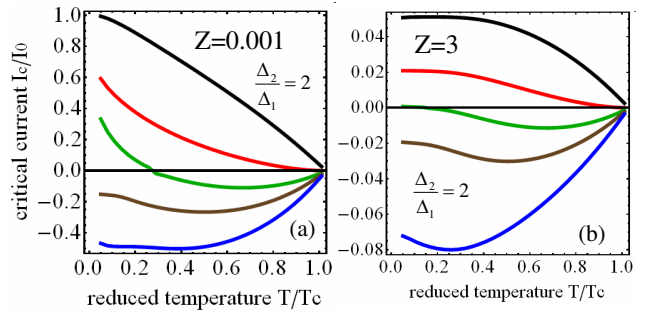}
 \caption{ Critical current as a function of
temperature. The contribution of the continuous energy spectrum
states has been added. (a)$Z=0.001$;(b)$Z=3$; in both figures, (a)
and (b), the black, red, green, brown, blue lines correspond to the
curves with $\alpha=0.001,0.5,0.6,0.65,0.7$, respectively.}
\label{fig:iron4}
\end{figure}

The negative sign of $I_c$ (blue lines, Fig. \ref{fig:iron3} a) and
b)) indicates that the junction free energy, i.e. the quantity
$F(\varphi)={\Phi_0}/{2 \pi}\int_0^{\varphi}d\varphi I(\varphi)$,
has a minimum at $\varphi=\pi$  such that the ground state of the
junction is $\pi$ shifted. As discussed above, the $\pi \rightarrow 0$
 crossover occurring with decreasing temperature for
$\alpha=0.65$ in figure \ref{fig:iron3} (a) and for $\alpha=0.6$ (green line) in figure \ref{fig:iron3} (b)(see inset) is due to the
competition between the contribution to the current from the
high energy band $E_2$ and that low energy band $E_1$.

 So far, by using Eq. (\ref{eq:currlev}), we have included
the contribution to the Josephson current from the discrete energy
spectrum only. However the continuous-spectrum states make their own
contribution to the current which has to be accounted for properly.
Following the approach developed by Furusaki Tsukada \cite{furusaki}
 the total Josephson current, including contributions from both the Andreev bound
states and the continous spectrum is given by
\begin{equation}
I=\frac{e \Delta_1k_B T}{\hslash}\sum_{\omega_n}\frac{1}{\Omega_1}\left[a(\varphi,i\omega_n,\alpha)-a(-\varphi,i\omega_n,\alpha)\right]
\end{equation}
Here we have defined $\Omega_1=\sqrt{\omega_n^2+\Delta_1^2}$ and introduced the Matsubara frequencies $\omega_n= \pi k_BT
(2n+1)$, with $n$ ranging from $-\infty$ to $+\infty$. $a(\varphi,i\omega_n,\alpha)$ is a scattering amplitude coefficient
for the process in which an electron-like quasiparticle traveling
from the left of the junction is reflected back as a hole-like
quasiparticle in the presence of two bands. This coefficient is derived by
 solving the BdG equations under the assumption $E>\Delta_1$
 and dropping the requirement of  exponentially decay of the
solutions for $|x|\rightarrow \infty$. The details of this procedure will be given elsewhere
\cite{nappi}.
 The
results of the calculated total critical current $I_c$
as a function of the reduced temperature, are shown in Fig.
\ref{fig:iron4}, (a) and (b), for different values of the band ratio
$\alpha$ and already considered in Fig.
\ref{fig:iron3}.

The general features of the curves $I_c$  vs $T$, shown in Fig.
(\ref{fig:iron4}) qualitatively reproduce  those of Fig.
(\ref{fig:iron3}). Most notably they confirm the $\pi-0$ junction
crossover. However some noticeable differences may be pointed out. For instance, the $\pi-
0$ step-like crossover, calculated for $\alpha=0.65$  and shown in
Fig. (\ref{fig:iron4}) (a), is smoothed out when considering the
continuous spectrum contribution, as can be seen by comparing with
the corresponding curve in Fig. (\ref{fig:iron3}), (b).

 In conclusion the  model investigated in this paper for a multiband superconductor
symmetric $s^\pm I s^\pm$ junction,  predicts a number of non
trivial details related to the Josephson effect in these systems.
These results are of relevant interest for the case of all-pnictide
Josephson micro-junctions. The spectral equation  provides the
Andreev bound levels as a function of the band ratio parameter
$\alpha$.  The main effect of the presence of the second band is the
building up of two extra Andreev levels which drive Cooper pairs in
a direction opposite to that observed in the presence of a single
band.

The phase-current relations predicted on the basis of the Andreev
levels (without consideration for the continuous energy spectrum)
shows  the formation of a noteworthy $\pi$-state in the junction
coupling. For selected values of the band ratio, a $\pi \rightarrow 0 $
crossover may occur as the temperature decreases.
In this case the $\pi$-state is observed at high temperature whereas the $0$-state
is observed below the crossover temperature. In this region the junction recovers
 the behavior of a conventional $''0''$ junction. These results are
 confirmed by means of a more exhaustive  evaluation of the Josephson current
 englobing the contribution of the continuous spectrum energy
 states.

\begin{acknowledgments}
We thank A. A. Golubov for illuminating discussions on this topic.
The financial contribution of EU NMP.2011.2.2-6 IRONSEA project nr.
283141 is gratefully acknowledged.
\end{acknowledgments}


\begin{thebibliography}{apssample}
\bibitem{note0} When $D>(1/r+1)/2$, this expression has to be limited to
$0<\varphi<\cos^{-1}\left[1-(r+1)/Dr\right]$ and $2\pi -\cos^{-1}\left[1-(r+1)/Dr\right]<\varphi<2\pi$.
\end{thebibliography}
\end{document}